\documentstyle[preprint,tighten,aps]{revtex}
\newcommand\eqn[1]{\begin{equation}#1\end{equation}}

\begin{document}
\title{Analysis of the superdefomed rotational bands} 
\author{G. A. Lalazissis and K. Hara \\
Physik-Department, Technische Universit\"at M\"unchen \\
D-85747 Garching bei M\"unchen, Germany}
\maketitle
\begin{abstract}
All available experimental data for the $\Delta I=2$ transition energies
in superdeformed bands are analyzed by using a new one-point formula.
The existence of deviations from the smooth behavior is confirmed in
many bands. However, we stress that one cannot necessarily speak about
staggering patterns as they are mostly irregular. Simulations of the
experimental data suggest that the irregularities may stem from the
presence of irregular kinks in the rotational spectra. This could be a
clue but, at the moment, where such kinks come from is an open question.

\end{abstract}
\newpage

A regular $\Delta I=2$ staggering pattern, in which the states differing
by four angular momentum units are shifted by a similar amount of energy,
found for the first time in the measured $\Delta I=2$ transition
energies of the yrast superdeformed (SD) band of the nucleus $^{149}$Gd
\cite{gd149} attracted much interest. Since then, it is reported that
similar effects have been observed for several SD bands in the nuclei of
the mass regions A$\sim$130 ($^{131-133}$Ce \cite{ce132}), A$\sim$150
($^{147-148}$Gd \cite{gd147,gd148b,gd148l}, $^{148}$Eu \cite{gd148b},
$^{154}$Er \cite{er154}) and A$\sim$190 ($^{192}$Tl \cite{tl192},
$^{191-194}$Hg \cite{hg191,hg194o,hg194n}, $^{195}$Pb \cite{pb195}).
>From the theoretical point of view, a lot of effort has been devoted to
understand this phenomenon \cite{ham,pav,mik,sun}. However, no definite
and decisive conclusion has been obtained till now and many questions
are still open.

Usually, the experimental data is presented using multi-point
formulae as for example the five-point formula \cite{gd149}
\eqn{ 
\Delta E_\gamma(I) \equiv {1 \over {16}} \big[6E_\gamma(I)
-4E_\gamma(I+2)-4E_\gamma(I-2)+E_\gamma(I+4)+E_\gamma(I-4) \big]}
This kind of presentation leads to a certain regularity (staggering)
in some of the observed bands. It is questionable, however, whether the
produced staggering does really exist in the spectrum. It should be
noted that such a multi-point formula will produce different patterns
depending on its multiplicity. This makes the analysis rather
formula-dependent and it could possibly lead to misintepretation of the
experimental data. 

In a recent paper \cite{hl96}, we have called the attention to this
point and proposed an alternative way of presenting the experimental
data by introducing the so-called ``one-point formula''. This formula is
free from the ambiguities which a multi-point formula may introduce. In
the present work, a new extended form of the one-point formula is
proposed and applied to the systematic analysis of all available
experimental data. The necessity of such an extension has been already
suggested in our earlier work \cite{hl96}. The new formula is capable of
handling a higher order global behavior (presence of a term $I^n$ with
$n > 3$) which indeed exists in the observed data.

The basic idea behind the one-point formula is to look at the deviation
of the $\Delta I$ =2 transition energy
\eqn{
\Delta E(I) \equiv E(I) - E(I-2)}
from its smoothly increasing part. After all, this was also the reason
of using a multi-point formula to remove the smooth lower order spin
dependence (i.e. lower order powers in the spin variable $I$) by
evaluating a derivative using finite differences.

Instead of taking derivatives, we subtract a polynomial of order $N$
in $I$ from $\Delta E(I)$ and define what we call the $N$th order
one-point formula
\eqn{
\Delta^{(1)}_{N} E(I) \equiv \Delta E(I) - Q_N(I),
~~~Q_N(I) = \sum_{m=0}^N q_m I^m
\label{2.1}}
where the coefficients $q_m$ are determined by minimizing the quantity
\eqn{
\chi(q_0,\cdots,q_N) \equiv \left[\Delta^{(1)}_{N} E(I)\right]^2
\label{2.2}}
with respect to $q_m$ (${\partial \chi \over {\partial q_m}}=0$). This
leads to a set of $N+1$ equations ($m=0,1,\cdots,N$)
\eqn{
\sum_{n=0}^N S_{m n} q_n = T_m,
~~~S_{m n} \equiv \sum_I^{\Delta I=2} I^m I^n,
~~~T_m \equiv \sum_I^{\Delta I=2} I^m \Delta E(I).
\label{2.3}}
Our previous one-point formula corresponds to $N=3$ \cite{hl96}. We note
that the smooth part $Q_N(I)$ which we subtract from $\Delta E(I)$ is
nothing other than a polynomial of order $N$ determined by the
$\chi$-square fit to $\Delta E(I)$. However, in practice, this formula
cannot be used in the present form particularly when the order of the
polynomial $N$ is larger than 3 since the equation (\ref{2.3}) is highly
ill-conditioned and this is the reason why one needs an extension. Thus,
we want to transform it into another form.

First, let us note that the replacement $I \rightarrow I-I_0$ does not
change the fitting procedure since the shape of the polynomial $Q_N(I)$
is unchanged by a parallel translation. Thus, the origin of the spin
values can be shifted freely. Secondly, the spin variable can be scaled
too ($I \rightarrow aI$) since the order of the polynomial remains the
same. These properties can be used to rewrite the polynomial in a
different form.

Shifting and scaling the spin values can be achieved most generally by a
linear mapping $I=ax+b$. The increment of $x$ is thus $\Delta x={\Delta
I \over a}$ ($\Delta I = 2$). We will choose $a={I_{max}-I_{min} \over
2}$ and $b={I_{max}+I_{min} \over 2}$ so that the range of $x$ becomes
[-1,+1], where $x=-1~(+1)$ corresponds to $I=I_{min}~(I_{max})$. The
polynomial in question may thus be written in the form
\eqn{
Q_N(I)=\sum_{n=0}^N p_n P_n(x).
\label{2.4}}
Here, we use the Legendre polynomial $P_n(x)$ instead of $x^n$. The
reason will be explained below. The resulting set of equations is
similar to (\ref{2.3}) but $q_n$ is replaced by $p_n$ and $I^n$ ($I^m$)
by $P_n(x)$ ($P_m(x)$). This representation has an advantage that there
holds the relation
\eqn{
S_{m n} = \sum_{x=-1}^{+1} P_m(x) P_n(x) = 0
~~~ {\rm if} ~m+n={\rm odd}.
\label{2.5}}
It means that the whole set of equations splits into two independent
sets of equations of smaller dimensions, one with $m,n=even$ and the
other with $m,n=odd$:
\eqn{
\sum_{n=even~ or~ odd}^N S_{m n} p_n = T_m,
~~~S_{m n}\equiv\sum_{x=-1}^{+1} P_m(x) P_n(x),
~~~T_m\equiv\sum_{x=-1}^{+1} P_m(x) \Delta E(I).
\label{2.6}}
This set of equations determines the coefficients $p_n$ and accordingly
the polynomial (\ref{2.4}) which represents the smooth part of $\Delta E(I)$.

As mentioned before, the original set of equations (\ref{2.3}) is highly
ill-conditioned. It is indeed so ill-conditioned that even the double
precision algorithm is not free from the numerical instability caused by
large losses of accuracy if $N$ is greater than 3. This problem can be
avoided by shifting and scaling the spin values as presented above. In
fact, the situation improves slightly if one uses the power series $x^n$
thanks to a property analogous to (\ref{2.5}). Nevertheless, this does
not fully resolve the numerical instability. The reason lies basically
in the fact that the zeros of $x^n$ are multiple and are all
concentrated at $x=0$. In contrast, all zeros of $P_n(x)$ are simple and
never coincide with one another for different $n$'s. The use of the
Legendre polynomial $P_n(x)$ instead of the power series $x^n$ is thus
essential for the numerical reliability (both accuracy and stability). 

In the present work, all available experimental data are analyzed using a
7th order one-point formula ($N=7$). Before doing this, let us first
introduce the corresponding ``filtered'' one-point formula by setting
the quantity $\Delta^{(1)}_{N} E(I)$ to zero if its absolute value is
smaller than or equal to the corresponding errorbar. This formula is
quite useful in practice. By construction, it shows whether the
deviation of the transition energy $\Delta E(I)$ from its smooth part is
physically significant or not.

For the sake of comparison, we first show both the one-point formula and
the corresponding filtered one in Fig. 1 for some selected cases. The
filtered formula indeed shows clearly where deviations occur. For this
reason, we will use only this formula for the presentation of data in
what follows. In Figs. 2 -- 6, we present the results for all the
observed SD bands. To avoid possible confusion, we have used the same
notation to label the SD bands as in the original experimental papers.
Since the spin assignment is not known in most of cases, our spin values
refer to $I-I_{ref}=2,4,\cdots$ where $I_{ref}$ is a unknown reference
spin. For the nucleus $^{192}$Tl, however, the reference spin assumes
the value $I_{ref}=0$ because spins are suggested by the experiment
\cite{tl192}.

It is seen that, in addition to the well known case of $^{149}$Gd, there
are also several other cases where the deviation from the smooth
behavior is beyond the experimental errorbars. However, we should like
to emphasize that one cannot necessarily speak about the ``regular
staggering'' patterns despite the presence of deviations as they are
mostly irregular. For $^{194}$Hg, we have used the data from a new
experiment \cite{hg194n}, but we observed no significant difference
between the new and old data \cite{hg194o}. It is also interesting to
note that there are two different experiments for the same band. Namely,
for the SD band 6 of $^{148}$Gd, we have the measurements from Berkley
with the gammasphere \cite{gd148b} and from Legnaro using the GASP
spectrometer \cite{gd148l}. Our results are shown in Fig. 3, see the
middle diagram of the upper- and lowermost panels, respectively.
However, one finds no similarity between these two diagrams. In fact, a
close examination of these two data shows that they do not quite agree
with each other and that their differences are beyond errorbars. This is
a serious problem since one does not know which data one should adopt.
This example shows that it is worth remeasuring not only this nucleus
but also all others (as they have nothing to compare) at different
laboratories for the sake of confirmation of data.

Finally, we show that one can simulate the experimental data. Let us
take the SD band C of $^{192}$Tl. We device a simple minded `model' by
assuming a smooth rotational spectrum $E(I)= AI(I+1)$, which has two
kinks created by shifting the energy `down' at I=16 and `up' at I=32 by
certain amount. In Fig. 7, we compare the experimental data (left side)
and the one obtained by such a ``theory'' (right side). One sees that
the basic feature is quite well reproduced. This suggests that the
presence of irregular kinks in the spectra may be responsible for the
appearance of the irregularity. For more complicated cases, we believe
that the patterns can also be simulated by similar models with more
kinks (not necessarily regular ones). As a matter of fact, such a
simulation has been recently done for diatomic molecular rotational
spectra \cite{hl97}.

To summarize, we have analyzed all available experimental data in terms
of an extended one-point formula and found many irregularities in the
measured $\Delta I=2$ transition energies. While the physics behind
this phenomenon is still unclear, one thing is certain: We cannot
necessarily speak about the presence of regular staggering. We have also
pointed out that it may be worth reconfirming all data at different
laboratories. Finally, we have simulated the SD band C of the nucleus
$^{192}$Tl in terms of a simple minded model, which seems to suggest
that the observed irregularities may stem from the presence of irregular
kinks in the spectrum. This may imply that a more careful estimate of
the experimental errorbars would be necessary.

\newpage
\parindent=2.5truecm
\centerline{\large FIGURE CAPTIONS}
\begin{description}

\item[Fig. 1] Seventh order one-point formula (left side) and the
corresponding filtered formula (right side) applied to some SD bands

\item[Fig. 2] Seventh order filtered one-point formula applied to 
SD bands of $^{131-133}$Ce and $^{147}$Gd

\item[Fig. 3] Seventh order filtered one-point formula applied to 
SD bands of $^{147-148}$Gd and $^{148}$Eu

\item[Fig. 4] Seventh order filtered one-point formula applied to 
SD bands of $^{149}$Gd, $^{154}$Er and $^{191}$Hg

\item[Fig. 5] Seventh order filtered one-point formula applied to 
SD bands of $^{191-194}$Hg and $^{192}$Tl

\item[Fig. 6] Seventh order filtered one-point formula applied to 
SD bands of $^{195}$Pb

\item[Fig. 7] Comparison between the experiment and the ``kink theory''
for the SD band C of $^{192}$Tl, see the text

\end{description}

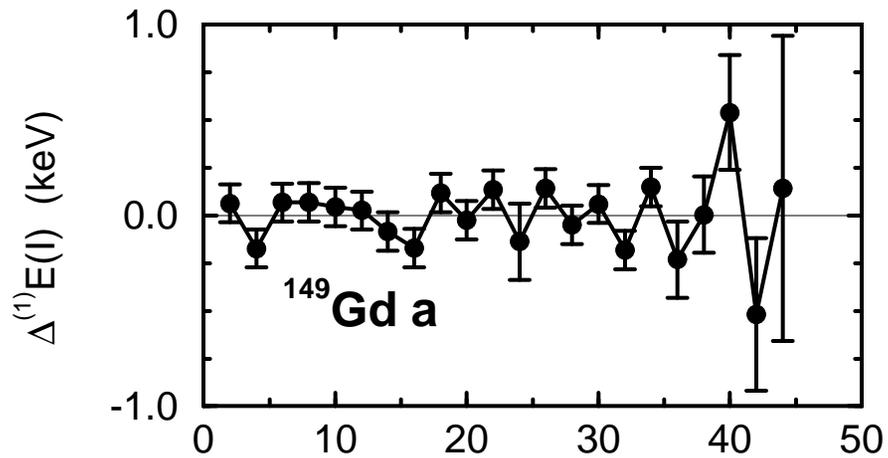
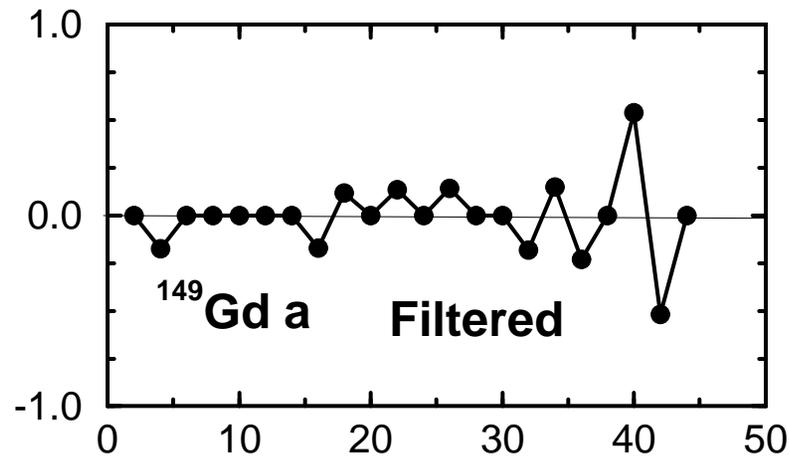
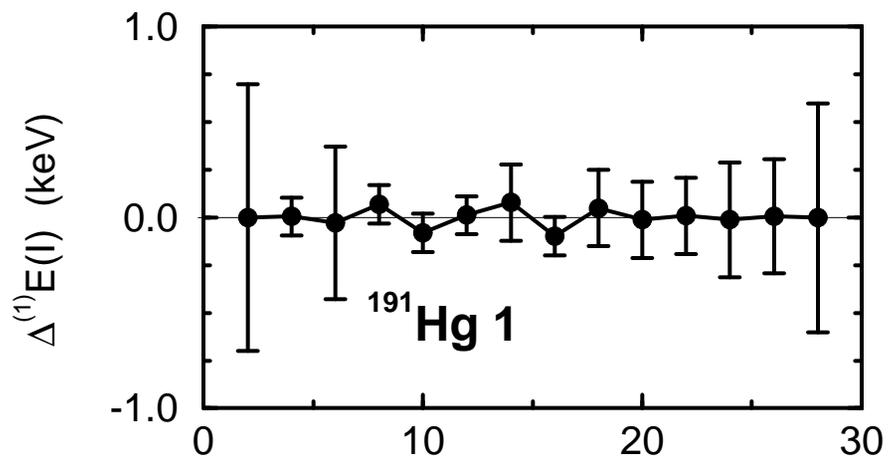
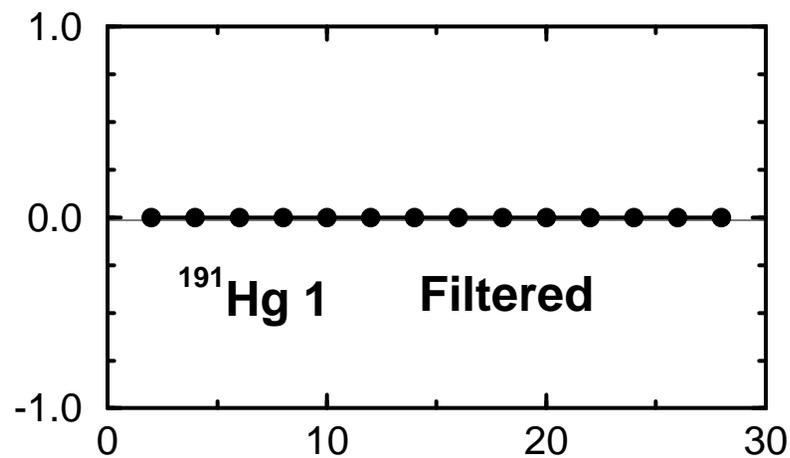
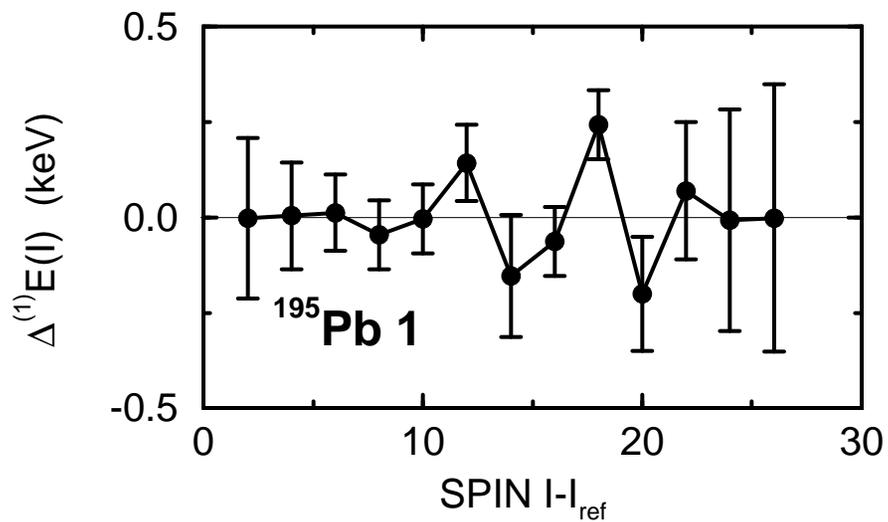
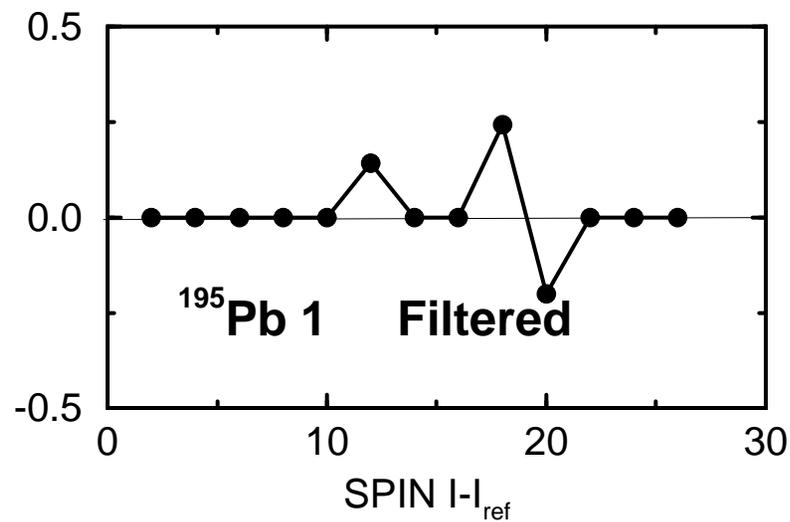
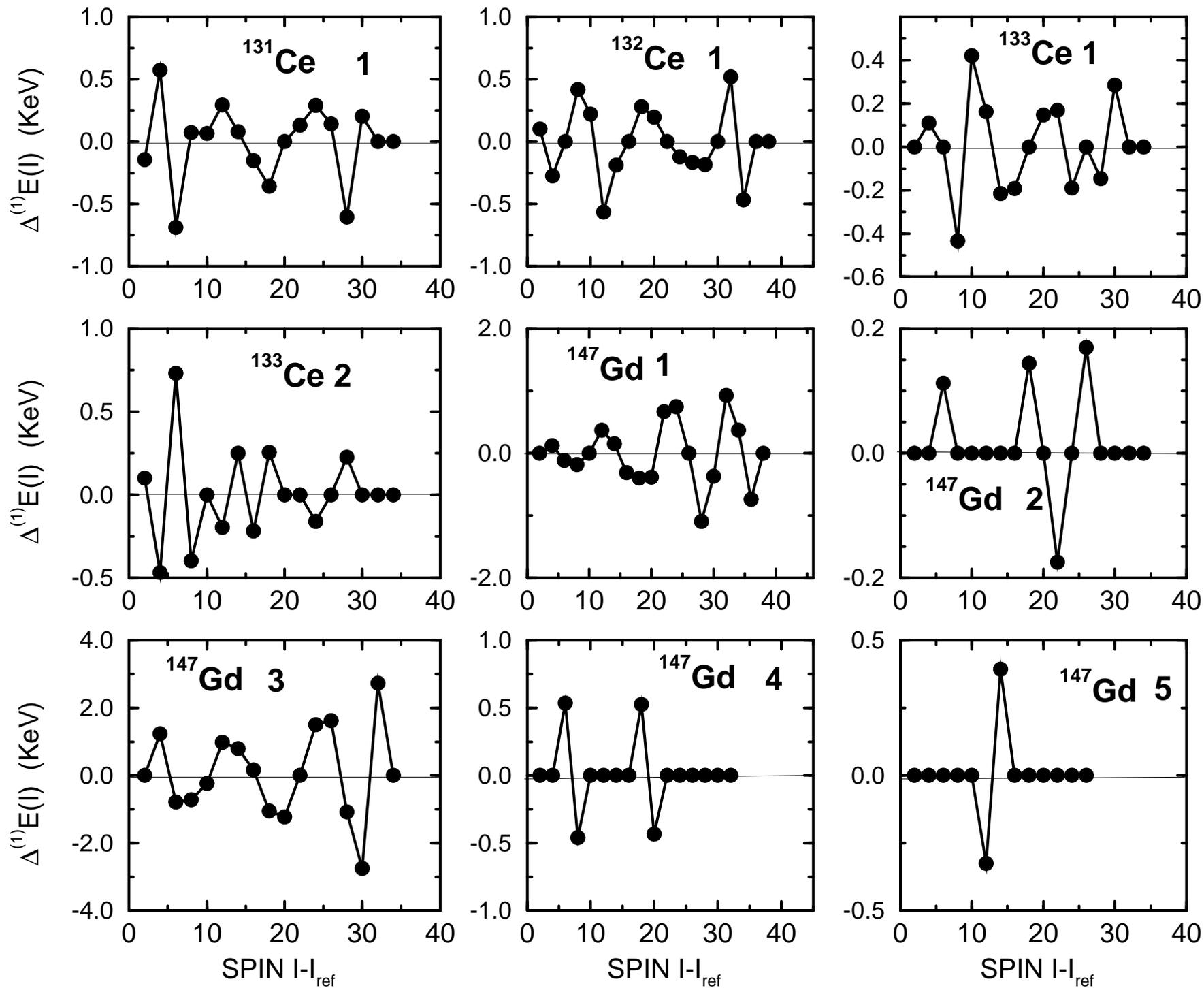
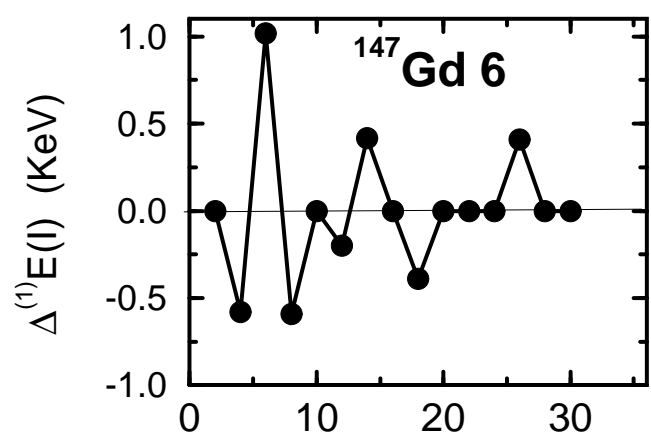
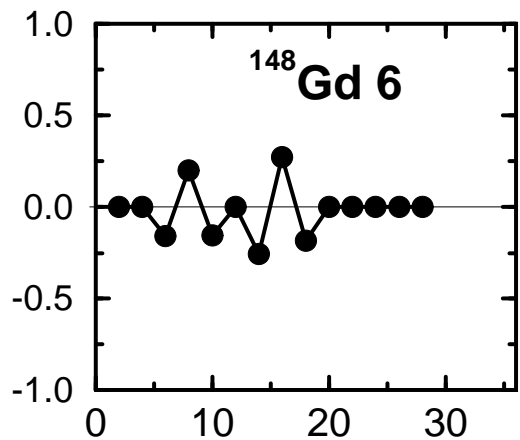
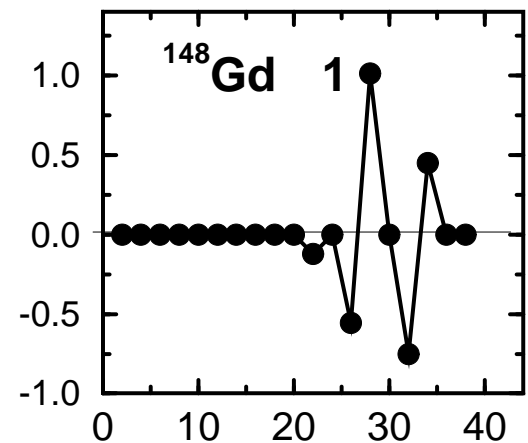
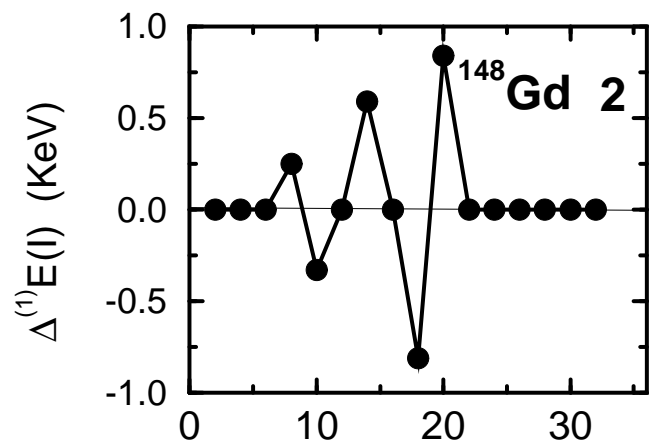
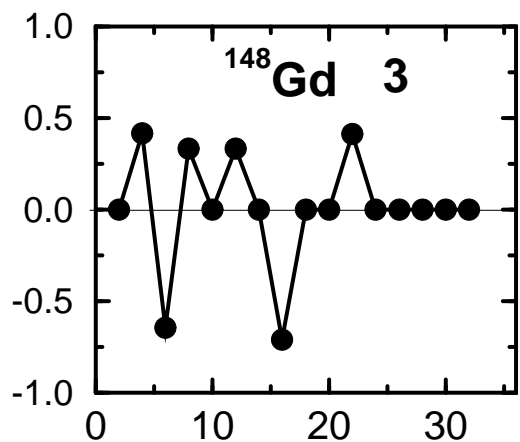
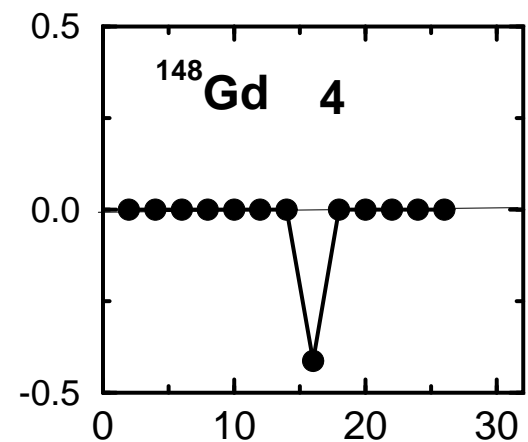
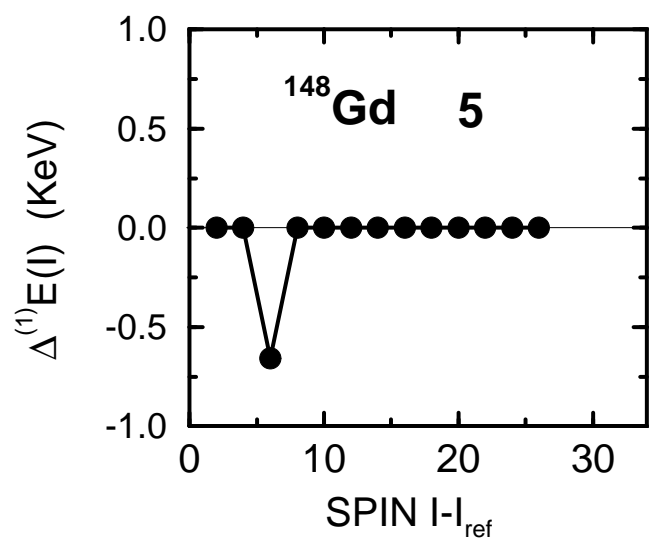
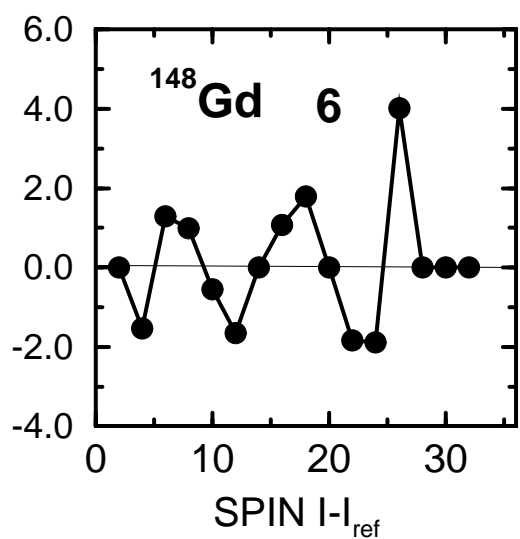
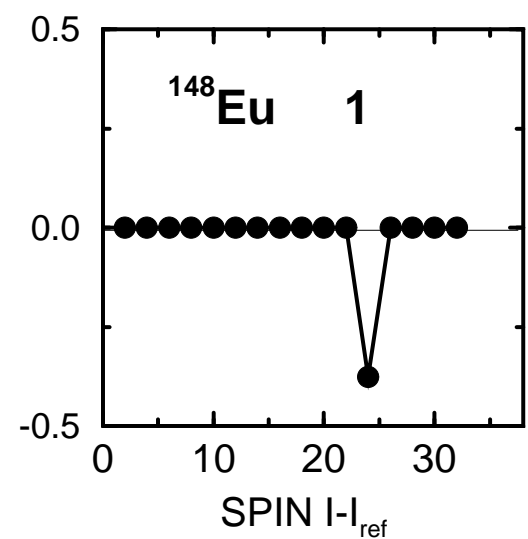
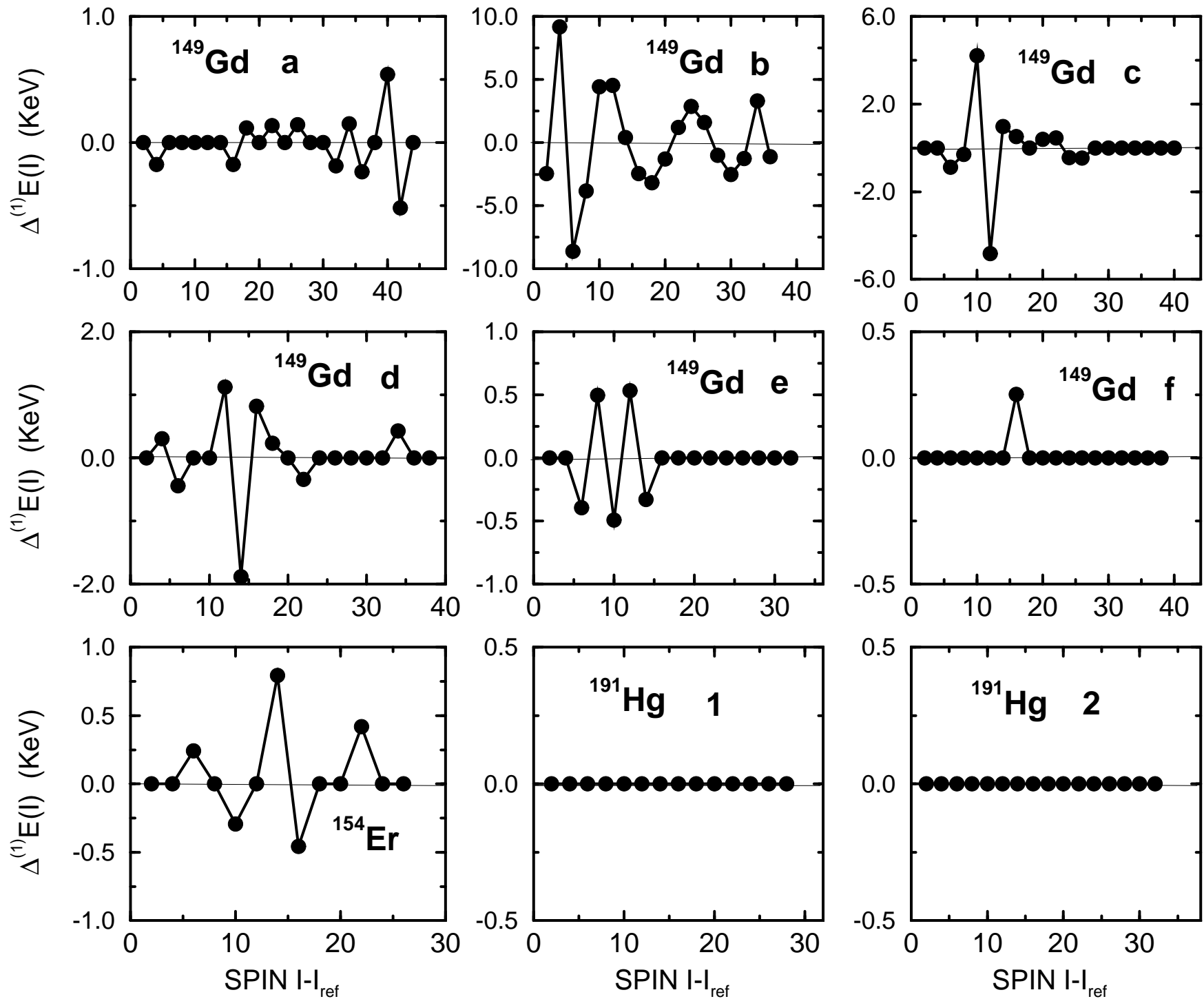
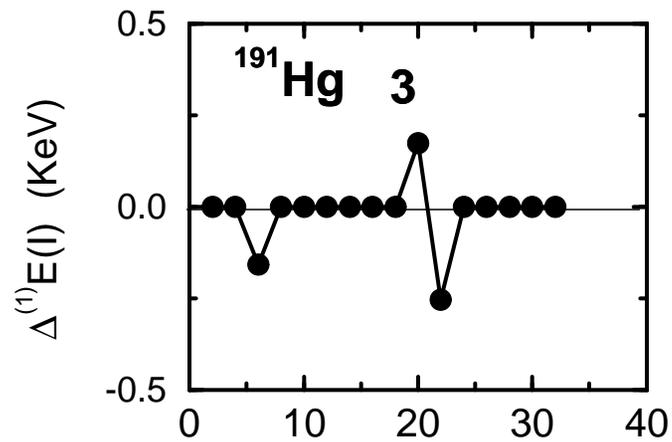
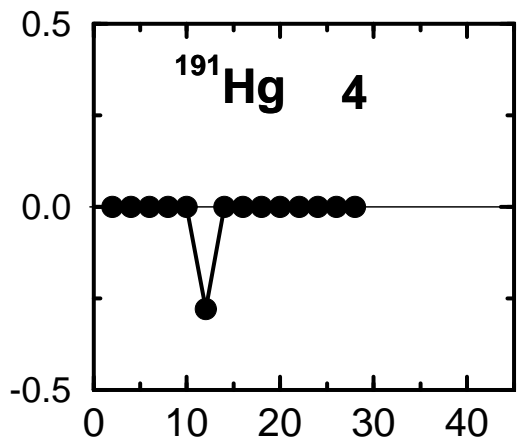
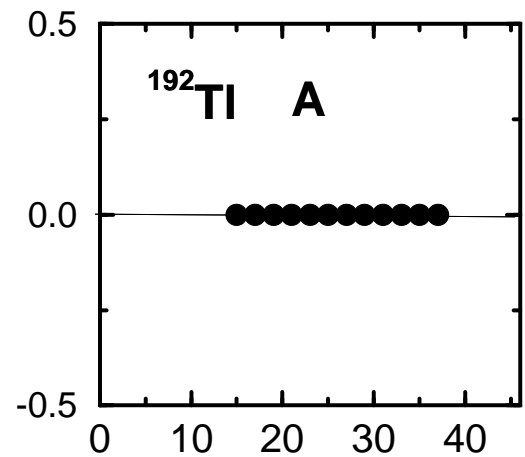
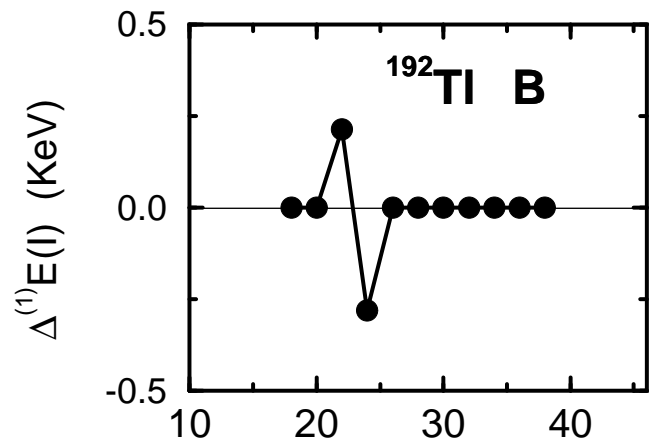
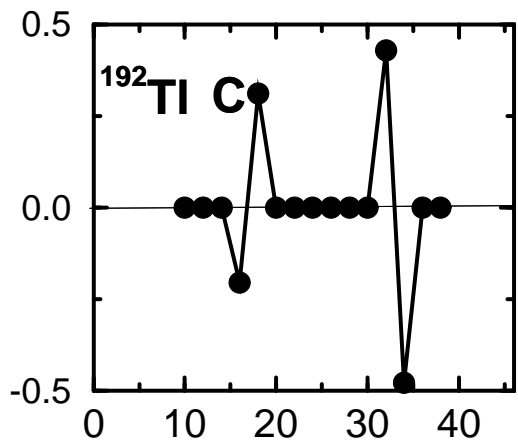
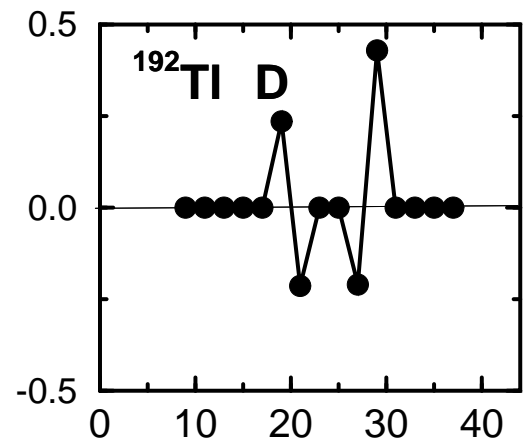
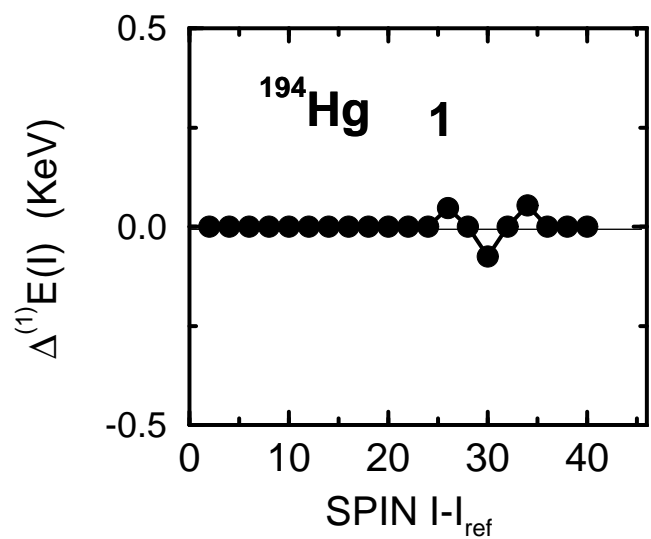
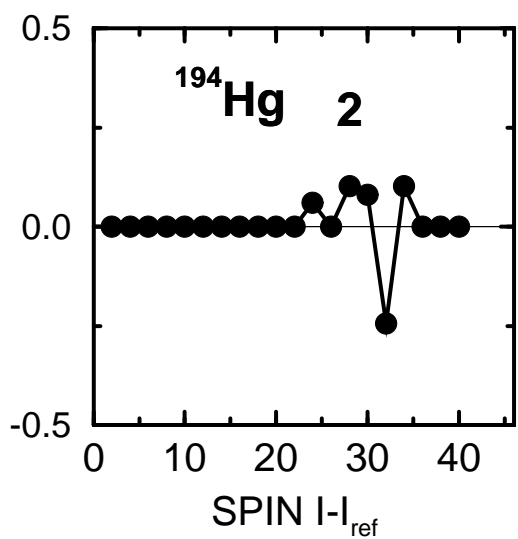
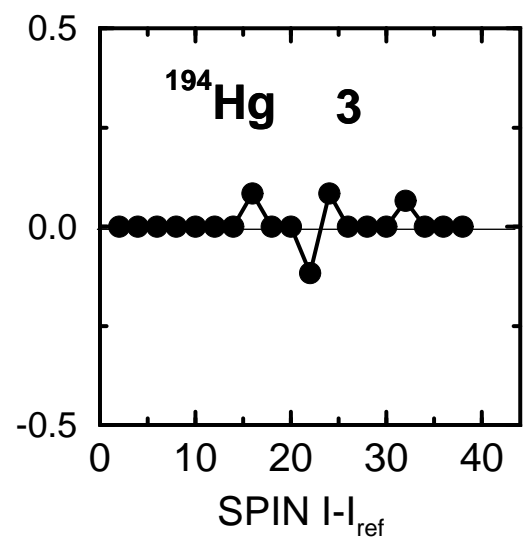
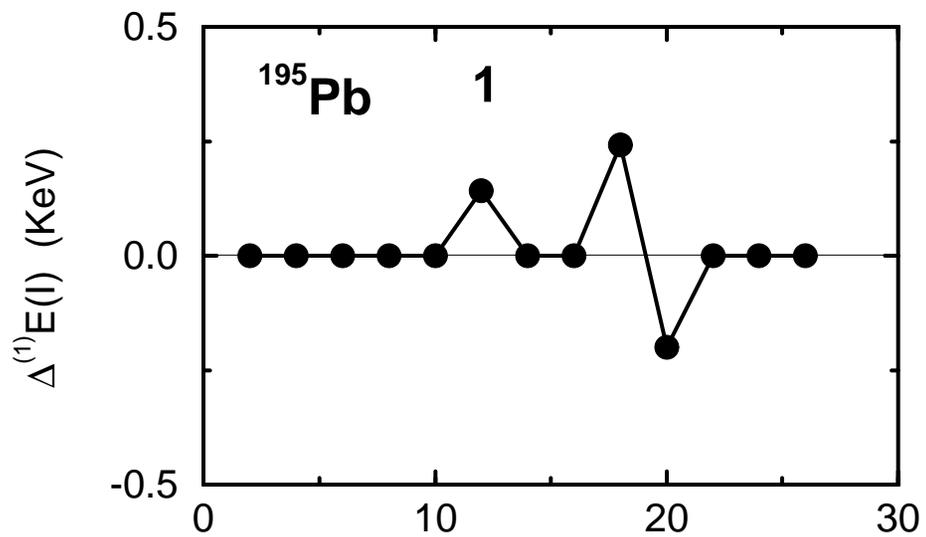
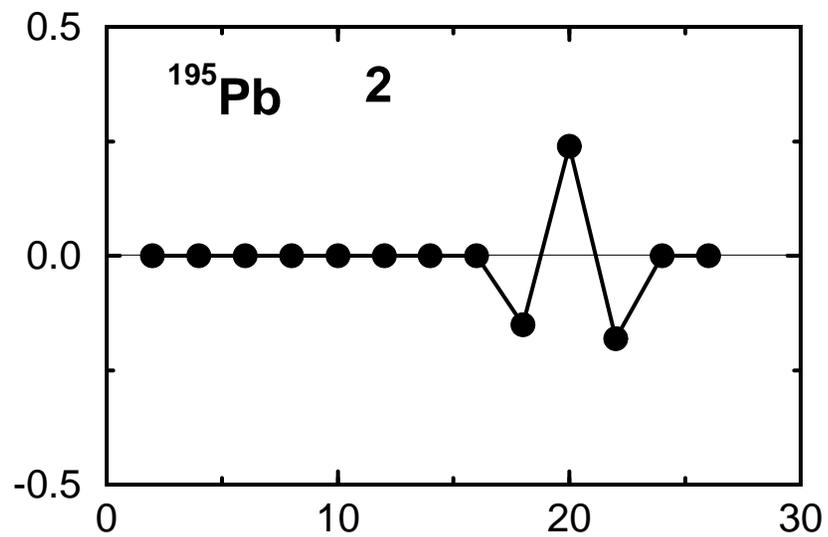
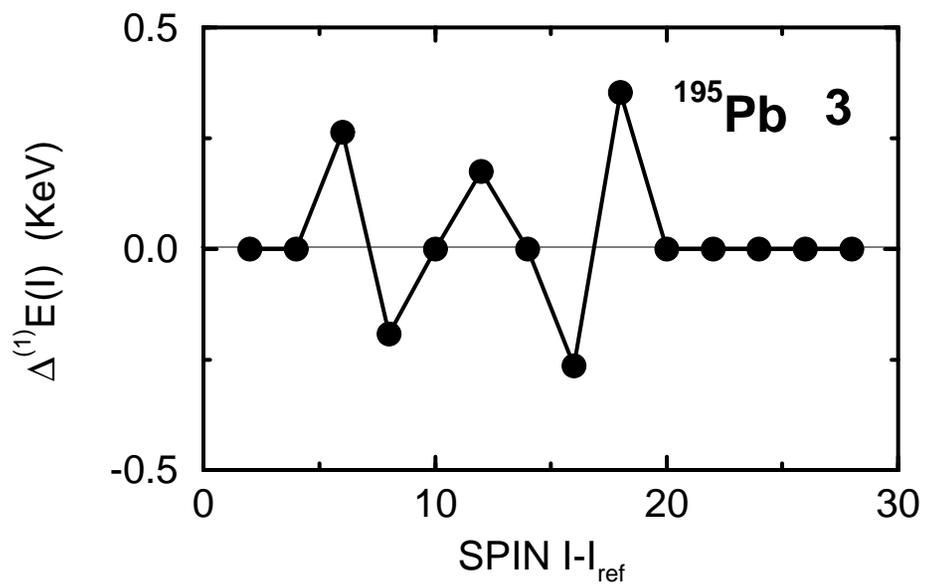
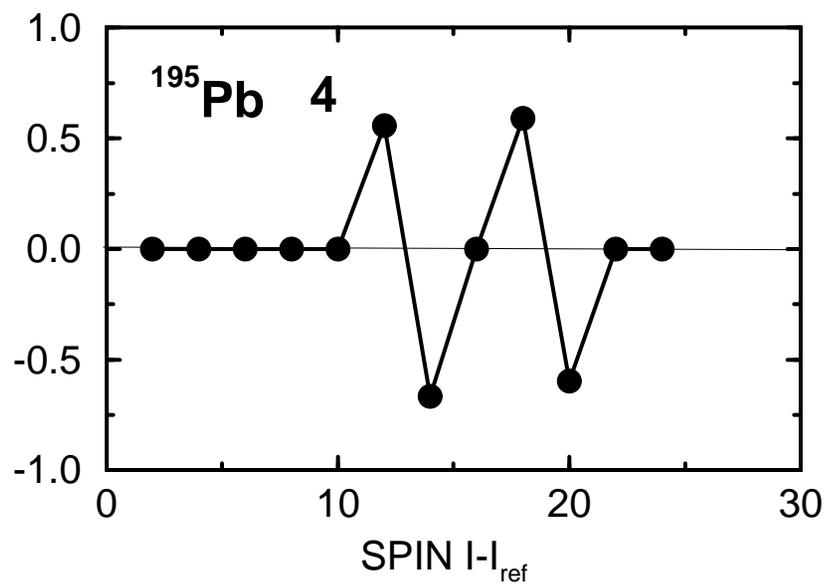
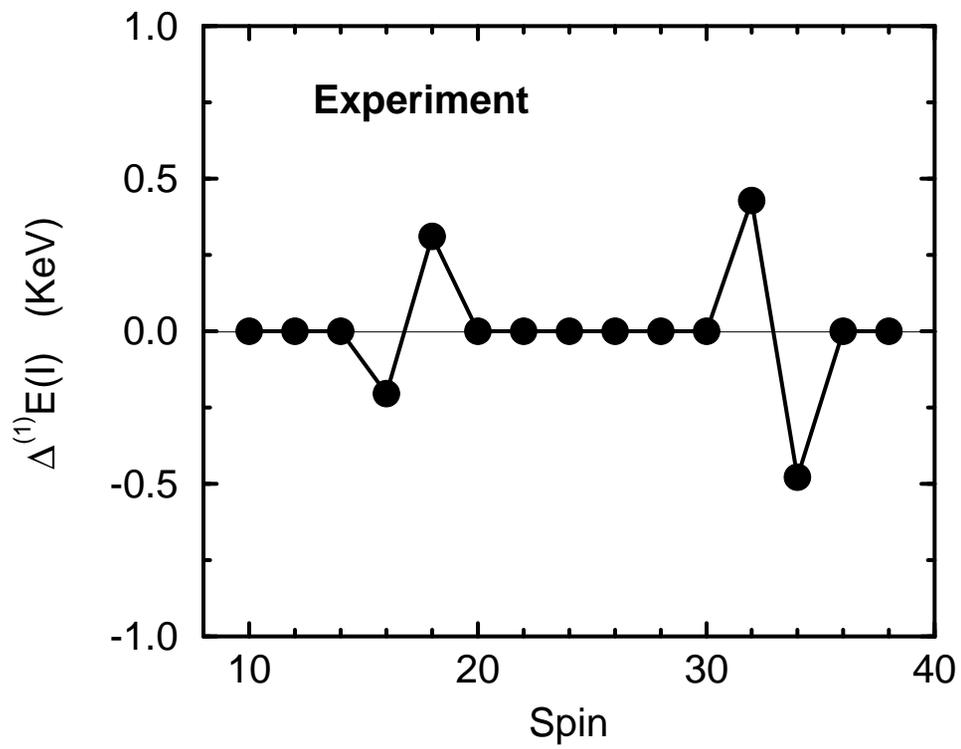
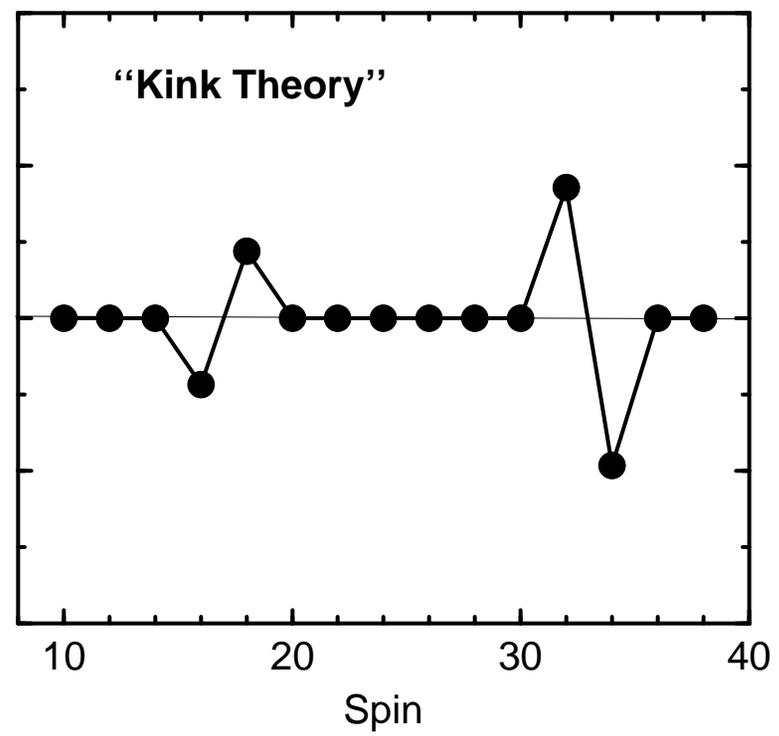
\end{document}